\documentclass[]{article}

\usepackage{caption}
\usepackage{subcaption}

\usepackage[nolist,nohyperlinks]{acronym}
\usepackage{amsmath,amssymb,amsfonts}
\usepackage{physics}
\usepackage{placeins}
\usepackage{float}
\usepackage{algpseudocode}
\usepackage{cleveref}
\usepackage{authblk}
\usepackage{graphicx} 
\usepackage{url} 
\usepackage{makecell}
\usepackage[table]{xcolor}

\algnewcommand\algorithmicforeach{\textbf{for each}}
\algdef{S}[FOR]{ForEach}[1]{\algorithmicforeach\ #1\ \algorithmicdo}

\begin{acronym}
    \acro{AI}{Artificial Intelligence}
    \acro{RS}{Recommender System}
    \acro{GRS}{Group Recommender System}
    \acro{GR}{Group Recommendation}
    \acro{DL}{Deep Learning}
    \acro{CF}{Collaborative Filtering}
    \acro{MF}{Matrix Factorization}
    \acro{MAE}{Mean Absolute Error}
    \acro{NDCG}{Normalized Discounted Cumulative Gain}
    \acro{BeMF}{Bernoulli Matrix Factorization}
    \acro{KNN}{K Nearest Neighbors}
    \acro{PMF}{Probabilistic Matrix Factorization}
    \acro{NMF}{non-Negative Matrix Factorization}
    \acro{BNMF}{Bayesian non-Negative Matrix Factorization}
    \acro{URP}{User Ratings Profile Model}
    \acro{NN}{Neural Network}
    \acro{MLP}{Multilayer Perceptron}
    \acro{NCF}{Neural Collaborative Filtering}
    \acro{IPA}{Individual Preference Aggregation}
    \acro{GPA}{Group Preference Aggregation}
    \acro{GMF}{Generalized Matrix Factorization}
    \acro{MLP}{Multi-Layer Perceptron}
    \acro{NDCG}{Normalizaed Discounted Cumulative Gain}
\end{acronym}
\acrodefplural{RS}[RSs]{Recommender Systems}

\begin{document}

\title{Neural Group Recommendation Based on a Probabilistic Semantic Aggregation}

\author[1,3]{Jorge Dueñas-Lerín}
\author[2,3]{Raúl Lara-Cabrera}
\author[2,3]{Fernando Ortega}
\author[2,3]{Jesús Bobadilla}

\affil[1]{Universidad Politecnica de Madrid, Madrid, Spain}
\affil[2]{Departamento de Sistemas Informaticos, Universidad Politecnica de Madrid, Madrid, Spain}
\affil[3]{KNODIS Research Group, Universidad Politecnica de Madrid, Madrid, Spain}

\date{}

\maketitle

\abstract{Recommendation to groups of users is a challenging subfield of recommendation systems. Its key concept is how and where to make the aggregation of each set of user information into an individual entity, such as a ranked recommendation list, a virtual user, or a multi-hot input vector encoding. This paper proposes an innovative strategy where aggregation is made in the multi-hot vector that feeds the neural network model. The aggregation provides a probabilistic semantic, and the resulting input vectors feed a model that is able to conveniently generalize the group recommendation from the individual predictions. Furthermore, using the proposed architecture, group recommendations can be obtained by simply feedforwarding the pre-trained model with individual ratings; that is, without the need to obtain datasets containing group of user information, and without the need of running two separate trainings (individual and group). This approach also avoids maintaining two different models to support both individual and group learning. Experiments have tested the proposed architecture using three representative collaborative filtering datasets and a series of baselines; results show suitable accuracy improvements compared to the state-of-the-art.}


\section{Introduction}\label{sec:introduction}

Personalization is one of the fields of \ac{AI} that has a greater impact on the lives of individuals. We can find a multitude of services that provide us with a personalized choice of news, videos, songs, restaurants, clothes, travels, etc. The most relevant tech companies make extensive use of personalization services: Amazon, Netflix, Spotify, TripAdvisor, Google, TikTok, etc. These companies generate their personalized recommendations using \ac{RS}~\cite{Batmaz2019} applications. \ac{RS} provide to their users personalized products or services (items) by filtering the most relevant information regarding the logs of items consumed by the users, the time and place that took place, as well as the existing information about users, their social networks, and the content of items (texts, pictures, videos, etc.). We can classify \acp{RS} attending to their filtering strategy as demographic~\cite{Bobadilla2021}, content-based~\cite{Deldjoo2020}, context-aware~\cite{Kulkarni2020}, social~\cite{Shokeen2020}, \ac{CF}~\cite{bobadilla2020dlarch, Dara2020} and filtering ensembles~\cite{Forouzandeh2021, cano2017hybridreview}. Currently, the \ac{MF}~\cite{salakhutdinov2007pmf} machine learning model is used to obtain accurate and fast recommendations between input data (votes). \ac{MF} translates the very sparse and huge matrix of discrete votes (from users to items) into two dense and relatively small matrices of real values. One of the matrices contains the set of short and dense vectors representing users, whereas the second matrix vectors represent items. Each vector element (real value) is called the `\emph{hidden factor value}’, since it represents some complex and unknown relationship between the input data (votes).

Whereas \ac{MF} machine learning models are fast and accurate, they also present a remarkable drawback: they cannot detect, in their hidden factors, the complex non-linear relationships between the input data. \ac{NN} can solve this problem through their non-linear activation functions. \ac{NN} based \ac{RS}~\cite{Bobadilla2022, Bobadilla2021} make a compression of information by coding the patterns of the rating matrix in their embeddings and hidden layers~\cite{Huang2020NeuralEC}. These embeddings play the role of the \ac{MF} hidden factors, enriching the result by incorporating non-linear relations. The most well-known \ac{NN} base \ac{RS} approaches are \ac{GMF} and \ac{MLP}~\cite{xiangnan2017ncf}.

\ac{GR}~\cite{ortega2013grs, Dara2020} is a subfield of the \ac{RS} area where recommendations are made to sets of users instead of to individual users (e.g.: to recommend a movie to a group of three friends). As in the regular \ac{RS}, the goal is to make accurate recommendations to the group. In this case, several policies can be followed; the most popular are: a) to minimize the mean accuracy error: to recommend the items that, on average, most like to all the group members, and b) to minimize the maximum accuracy error: to recommend the items that does not excessively dislike to any of the group members. It is important to state that there are not open datasets containing group information to be used by group recommendation models; for this reason, generally randomly generated groups are used for training and testing research models. 

Regardless of the machine learning approach used to implement \ac{GR} \ac{RS}, the most notable design concept is to establish where to locate the aggregation stage to convert individual information to group information. The general rule is: the sooner the aggregation stage, the better the performance of \ac{GR}~\cite{ortega2013grs}. There are three different locations where group information can be aggregated into a unified group entity: a) before the model, b) in the model, and c) after the model. The most intuitive approach is to combine individual recommendations into a unified group recommendation (option c)~\cite{Baltrunas2010}. This approach is known as \ac{IPA} and requires processing several individual recommendations followed by rank aggregation. However, the process is slow and not particularly accurate. On the other hand, to consider the entire group for the recommendation, we should work before or inside the model (options a or b). These approaches are known as \ac{GPA}. Aggregating group information before the model requires working with the user-item interaction matrix in a higher-dimensional space, which can lead to misinformation problems. To aggregate group information in the model, we need to work with the user`s hidden vector in the low-dimensional space.

Aggregating several hidden vectors from individual users into a unified virtual user hidden vector~\cite{ortega2016groupsmfcf} avoids compute the model predictions several times and makes the rank aggregation stage unnecessary. In addition, it takes advantage of operating with condensed information coming from the \ac{MF} compression of information: the virtual user can be obtained simply by averaging the representative short and dense vectors of the users group; this is efficient and accurate. An interesting question is: Can \ac{NN} operate the same way that \ac{MF} does to obtain virtual users and generate recommendations? First, note that many \ac{NN} based \ac{RS} models compress the user embedding in a different latent space than the item embedding, and it can be a problem; then, the \ac{NN} non-linear ensemble representations are more complex than the \ac{MF} hidden factor representations; consequently, simply averaging the ensembles of the users in the group does not automatically ensure a representative virtual user embedding. Furthermore, model-based aggregations (option ‘b’ in the previous paragraph) are model dependent, and then it is necessary to design and test different solutions for different \ac{NN} based \ac{RS} models. Whereas the \ac{NN} latent spaces are the state-of-the-art to catch users and items relations, some other machine learning approaches have been designed, such as the use of the random walk with restart method~\cite{Feng2019} providing a framework to relate users, items, and groups, and to exploit the item content and the profiles of the users. A three-stage method~\cite{ABOLGHASEMI20221} is proposed to increase the precision and fairness of \ac{GR}, where binary \ac{MF}, graphs and the dynamic consensus model are processed sequentially. Some relevant and current GR research aims to make use of the concept of member preference (influence or expertise) concept, based on similarity and trust. The key idea is to detect the group leaders as group members that are trusted more than others and have more influence than others. In~\cite{BARZEGARNOZARI2020106296}, fuzzy clustering and an implicit trust metric are combined to find neighborhoods. \ac{GR} based on an average strategy applied to user preference differences~\cite{Wang2020} has been combined with trusted social networks to correct recommendations. An aggregation approach for GR mimics crowd-sourcing concepts to estimate the level of expertise of group members~\cite{ISMAILOGLU2022113663}; it is implemented using parameters of sensitivity and specificity. The impact of social factors on a \ac{GR} computational model is evaluated in~\cite{Guo2016}, using the expertise factor, the influences of personality, preference similarities, and interpersonal relationships.

In this paper, we present how we can generate \ac{GR} using \ac{NN} based \ac{RS} by training the model using raw \ac{CF} data (i.e., the ratings of individual users to the items without any additional information). The \ac{GR} have been tested using the two most popular implementations of \ac{NN} based \ac{RS}: \ac{GMF} and \ac{MLP}. As stated previously, to make recommendations to groups using \ac{NN} based \ac{RS}, information of the individual users must be aggregated. The chosen information aggregation design is to merge the users of the group in the input vector that feeds the user embedding of the \ac{NN}. This aggregation design is not novel, since it has been used by \cite{sajjadi2021deepgroup} applied to a \ac{MLP} architecture. However, our approach combines several innovative aspects in comparison to the state of the art. On the one hand, the aggregation of the users in the group is a probabilistic function rather than a simple multi-hot encoding~\cite{sajjadi2021deepgroup}; this better captures the relative importance of users in the input vector that feeds the \ac{NN}, moreover: this aggregation approach serves as front-end for any \ac{NN} \ac{GR} model. On the other hand, we propose the use of a simple \ac{RS} \ac{NN} model (\ac{GMF}) instead of the deepest \ac{MLP} one~\cite{sajjadi2021deepgroup}; the hypothesis is that complex models overfit \ac{GR} scenarios, since they are designed to accurately predict individual predictions, whereas \ac{GR} must satisfy an average of the tastes in the group of users, that is, \ac{GR} should be designed to generalize the set of individual tastes in the group. Furthermore, the proposed architecture just needs a single training to provide both individual recommendations and group recommendations; particularly, the model is trained by only using individual recommendations (as in regular \ac{RS}). Once the model is trained to return individual predictions, we can fill the input vector by aggregating all the users in the group, then feed-forward the trained model and finally obtain the recommendation for the group of users. Anyway, the impact of these innovative aspects can be evaluated in \cref{sec:experiments}, where we empirically compare the proposed aggregation designs with respect to the main baseline~\cite{sajjadi2021deepgroup}
 
In summary, the \ac{GR} state-of-art presents the following drawbacks: a) Some research relies on additional data to the \ac{CF} ratings, such as trust or reputation information that is not available on the majority of datasets, b) different proposals make the aggregation of individual users before (\ac{IPA}) or after (Ranking) the model, making it impossible to benefit from the machine learning model inner representations (\ac{GPA}), and c) The proposed GR neural model solutions tend to apply architectures designed to make individual recommendations, rather than group ones; this leads to the model overfitting and to a low scalability referred to the number of users in a group. To fill the gap, our proposed model: a) Acts exclusively on \ac{CF} ratings, b) Makes user aggregation in the model, and c) Its model depth and design enables adequate learning generalizations. Additionally, the provided experiments test the proposed model according to different aggregation strategies to set the group labels used in the learning stage. In contrast, a notable limitation of our architecture and the experiments is the lack of testing on particularly demanding scenarios such as cold start in groups users, extremely sparse data sets, impact of popular item bias, and fear \ac{GR}.

The rest of the paper is structured as follows: \Cref{sec:model} introduces the tested models and aggregation functions; \Cref{sec:experiments} describes the experiment design, the selected quality measures, the chosen datasets and shows the results obtained; \Cref{sec:discussion} provides their explanations; and \Cref{sec:conclusions} highlights the main conclusions of the article and the suggested future work.

\section{Proposed model}\label{sec:model}

\begin{figure}
\centering
\begin{minipage}{.5\textwidth}
  \centering
  \includegraphics[width=.8\linewidth]{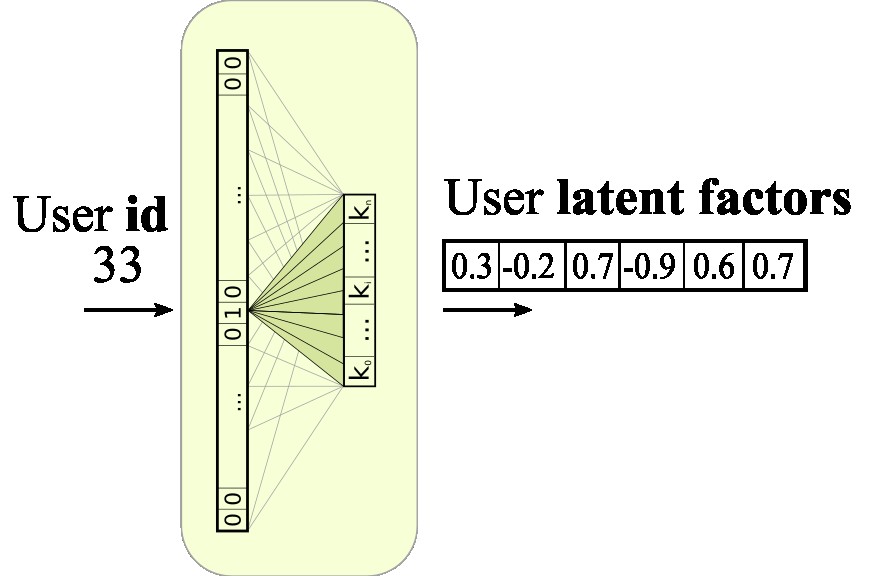}
  \captionof{figure}{Embedding layer schema.}
  \label{fig:embedding}
\end{minipage}%
\begin{minipage}{.5\textwidth}
  \centering
  \includegraphics[width=.8\linewidth]{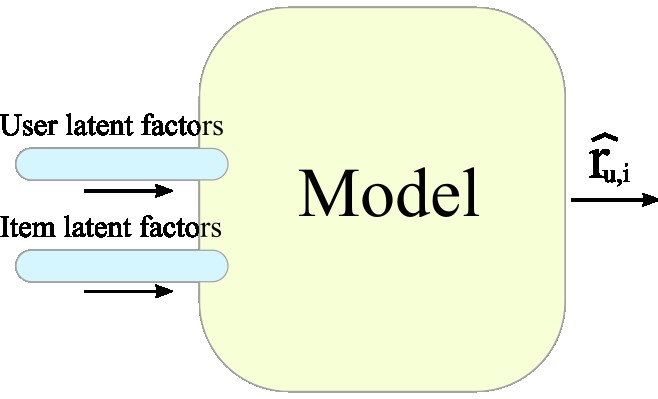}
  \captionof{figure}{Collaborative Filtering based Neural Network Model.}
  \label{fig:embedding-model}
\end{minipage}
\end{figure}

In \ac{CF} interactions (purchase, viewing, rating, etc.) between users and items are stored in a sparse matrix since it is common for users to interact only with a small proportion of the available items and, in the same way, only a small percentage of existing users interact with the items. The sparsity levels of this matrix is around 95-98\% as shown in \cref{tab:main_parameters}. To handle this sparsity, current \ac{CF} models based on \ac{NN}~\cite{xiangnan2017ncf} work with a projection of users and items into a low-dimensional latent space using Embedding layers. Embedding layers are a very popular type of layer used in \ac{NN} that receive as input any entity and return a vector with a low-dimensional representantion of the entity in a latent space. These vectors are commonly named \emph{latent factors}. In order to transform the entity into its low-dimensional representation, the embedding layer first transforms the entity into a one hot encoding representation (typically using a hash function). \Cref{fig:embedding} sumarizes this process.

In the context of \ac{CF}, two Embedding layers are required: one for the users and the other for the items. Later, both Embedding layers are combined using a \ac{NN} architecture (see~\Cref{fig:embedding-model}). For example, the aphormented models \ac{GMF} and \ac{MLP} uses a Dot layer and a Concatenate layer followed by some fully connected dense layers as architectures, respectively. 

Formally, we define a \ac{NN} model $\Phi$ that predicts the rating that a user $u$ will give to an item $i$ ($\hat{r}_{u,i}$) combining the latent factors provided by the Embedding layer ($Emb_L$) of the user $u$ ($\vec{l_u}$) and the the item $i$ ($\vec{l_i}$):

\begin{equation}
\begin{split}
    Emb_L(u) = \vec{l_u}\\
    Emb_L(i) = \vec{l_i},\\
    \Phi (\vec{l}_{u}, \vec{l}_{i}) = \hat{r}_{u,i}
\end{split}
\end{equation}

As stated in~\cref{sec:introduction}, when working with \ac{GR}, a straightforward strategy is \ac{IPA}~\cite{liang2014deepmodelinggrouppreferences}. This strategy makes a prediction for each member of the group and then performs an aggregation. This strategy does not treat the group as a whole. If we have a group of users $G = \{u_1, u_2, ..., u_n\}$, the prediction of the rating of this group $G$ to an item $i$ ($\hat{r}_{G,i}$) is computed as the average value of the individual predictions:

\begin{equation}
    \hat{r}_{G,i} = \dfrac{1}{{\|G\|}} \sum _{u \in G} \hat{r}_{u,i} = \dfrac{1}{{\|G\|}} \sum _{u \in G} \Phi(\vec{l_u}, \vec{l_i})
    \label{eq:fullgpa}
\end{equation}

On the other hand, the \ac{GPA} strategies take into account the group as a whole. It should be noted that the order of users within the group and the length of it should not affect the aggregation; thus, the aggregations should meet the constraints of: permutation invariant and fixed result length~\cite{sajjadi2021deepgroup}. Our goal with the \ac{GPA} strategy is to be able to obtain a prediction $\hat{r}_{G,i}$ with a single forward propagation and to treat the group as a whole entity. We can achieve this by aggregating the latent factors of each user that belongs to the group to obtain the latent factor of the group $\vec{l}_{G}$. Once the latent factors of the group are aggregated, the model $\Phi$ can be used to compute the predictions:

\begin{equation}
\begin{split}
    Emb_L(G) = \vec{l}_{G}\\
    \hat{r}_{Gi} = \Phi(\vec{l}_{G}, \vec{l}_{i})
\end{split}
\label{eq:grouplatent}
\end{equation}

The aggregation of group latent factors in embedding layers can be achieved by modifying the input of the \ac{NN}. As mentioned previously, Embedding layers have as input a one hot representation of the entities. This approach is adequate when performing individual predictions, however, for group recommendations, we need to apply a multi-hot representation to the users Embedding layer, i.e., we encode the group by setting multiple inputs of the user Embedding layer (the inputs related with the users that belong to the group) to a value higher than 0. This encoding allows us to take into account all group users at the same time for the extraction of latent factors of the group $\vec{l}_{G}$.

The simplest aggregation, which is used by the DeepGroup model~\cite{sajjadi2021deepgroup}, is to use as input for embedding a constant value proportional to the size of the group. We define the input of the user's Embedding layer for the user $u$ as

\begin{equation}
    EmbeddingInput_{Average}(u) = \begin{cases} 
        \frac{1}{{\|G\|}} & if u \in G\\
        0 & if  u \notin G
   \end{cases}\label{eq:input_average}
\end{equation}

We call this aggregation `\emph{Average}' since the embedding layer will generate the group latent factor equal to the average of the latent factors of all users in the group.

\ac{RS} can give better predictions the more information they have about users, so to take advantage of this fact, we have tested the `\emph{Expertise}' aggregation in which we give a weight to the users proportional to the number of votes they have entered into the system. Let ${{\|R_u\|}}$ the number of ratings of the user $u$, the input of the users' Embedding layer for the user $u$ is defined as

\begin{equation}
    EmbeddingInput_{Expertise}(u) = \begin{cases} 
        \frac{{\|R_u\|}}{\sum _{g \in G} {\|R_g\|}} & if u \in G\\
        0 & if  u \notin G
   \end{cases}\label{eq:expert}
\end{equation}

In addition to the `\emph{Expertise}' aggregation, we also proposed the `\emph{Softmax} aggregation as a smooth version of the `\emph{Expertise}' aggregation. In this case, the input of the users' Embedding layer for the user $u$ is defined as

\begin{equation}
    EmbeddingInput_{Softmax}(u) = \begin{cases} 
        \frac{e^{\|R_u\|}}{\sum _{g \in G} e^{\|R_g\|}} & if u \in G\\
        0 & if  u \notin G
   \end{cases}\label{eq:softmax}
\end{equation}

\begin{figure}[ht]
    \centering
    \includegraphics[width=0.8\textwidth]{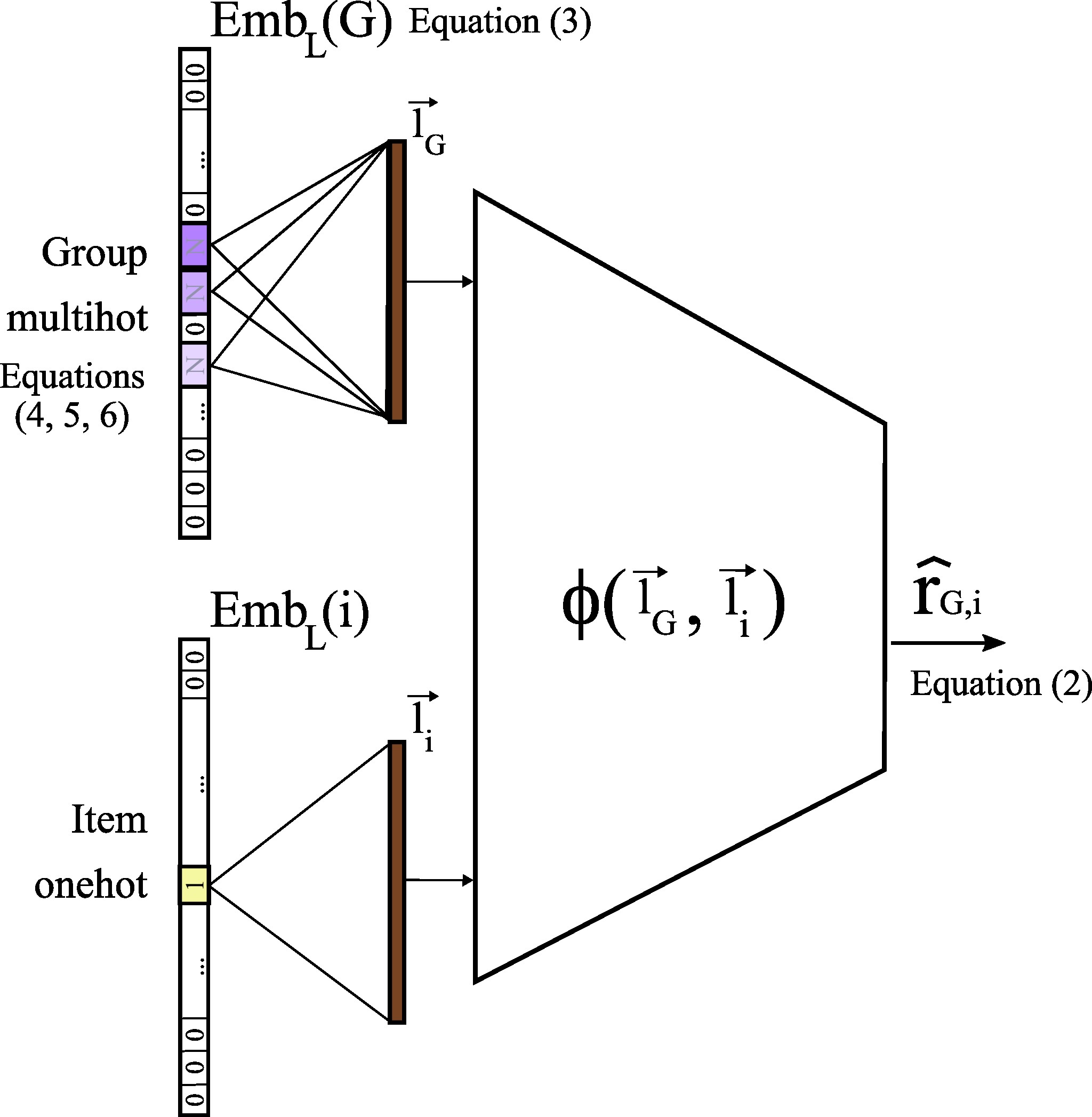}
    \caption{Graphical representation of the proposed model.}
    \label{fig:completeprocess}
\end{figure}

In \Cref{fig:completeprocess} we can see where the equations fit in the group recommendation process. The first step is to generate the multi-hot vector with some of the described aggregation (\cref{eq:input_average,eq:expert,eq:softmax}). This vector (multi-hot representation of the group) is fed into the embedding layer to obtain a vector of the latent factors of the groups $\vec{l}_{G}$ (\cref{eq:grouplatent}). Once the latent factors of the group and the item are obtained, they are used to feed the model $\Phi$ (\ac{GMF} or \ac{MLP}) and produce the rating prediction for the group $G$ on the item $i$ (\cref{eq:fullgpa}).

In \Cref{tab:inputexample} we can find an example with some users (13, 24, 30 and 42) with different rating counts (\Cref{tab:ratingcount}), their input values to the users' Embedding layer in a multi-hot fashion (\Cref{tab:input}), their individual latent factors (\Cref{tab:latentfactors}), and the final group latent factors with different aggregations (\Cref{tab:aggexample}).

\begin{table}[h]
    \centering
    \begin{subtable}[h]{0.9\textwidth}
        \centering
        \begin{tabular}{|c|c|c|c|c|c|c|c|c|c|}
        \hline
        User     & ... & $u_{13}$ & ... & $u_{24}$ & ... & $u_{30}$ & ... & $u_{42}$ & ... \\ \hline
        \#Rating &     & 2      &     & 5      &     & 6      &     & 3      &     \\ \hline
        \end{tabular}
       \caption{Rating count.}
       \label{tab:ratingcount}
    \end{subtable}
    \begin{subtable}[h]{0.9\textwidth}
        \centering
        \begin{tabular}{|c|c|c|c|c|c|c|c|c|c|}
\hline
Strategy\textbackslash{}User     & ... & $u_{13}$ & ... & $u_{24}$ & ... & $u_{30}$ & ... & $u_{42}$ & ... \\ \hline
Average                      &     & 0,25  &     & 0,25 &     & 0,25 &     & 0,25 & \\ \hline
Expertise                    &     & 0,13  &     & 0,31 &     & 0,38 &     & 0,19 & \\ \hline
Softmax                      &     & 0,01  &     & 0,26 &     & 0,70 &     & 0,03 & \\ \hline
\end{tabular}
        \caption{Input values to the users' Embedding layer.}
        \label{tab:input}
     \end{subtable}
     
    \begin{subtable}[h]{0.9\textwidth}
        \centering
\begin{tabular}{|c|c|c|c|}
\hline
User\textbackslash{}factor & $l_1$   & $l_2$   & $l_3$   \\ \hline
$u_{13}$ & 0,1 & 0,6 & 0,3 \\ \hline
$u_{24}$ & 0,7 & 0,2 & 0,9 \\ \hline
$u_{30}$ & 0,8 & 0,4 & 0,1 \\ \hline
$u_{42}$ & 0,5 & 0,7 & 0,8 \\ \hline
\end{tabular}
    \caption{Users' latent factors assuming a latent space of size 3.}
     \label{tab:latentfactors}
    \end{subtable}

    \begin{subtable}[h]{0.9\textwidth}
        \centering
\begin{tabular}{|c|c|c|c|}
\hline
Agg\textbackslash{}factor & $l_{G_1}$   & $l_{G_2}$   & $l_{G_3}$   \\ \hline
Average     & 0,525 & 0,475 & 0,525 \\ \hline
Expertise   & 0,629 & 0,425 & 0,508 \\ \hline
Softmax     & 0,758 & 0,359 & 0,331 \\ \hline
\end{tabular}
    \caption{Group latent factors using different aggregations.}
     \label{tab:aggexample}
     \end{subtable}

     \caption{Complete aggregation example.}
     \label{tab:inputexample}
\end{table}

\section{Experimental evaluation}\label{sec:experiments}

In this section, we show the experiments carried out to validate the aggregation proposed in this manuscript. As previously stated, the experiments have been performed using the most popular \ac{NN} based \ac{RS} architectures: \ac{GMF} and \ac{MLP}. We have chosen these two architectures because they are the best known and offer the best results for individual predictions. However, the aggregation strategies proposed can be applied to any \ac{NN} architecture based on Embedding layers.

The choice of datasets has been made considering that: a) there are no open datasets containing information on group voting; and b) \ac{GMF} and \ac{MLP} models should be trained using individual voting, since the proposed aggregations allow computing predictions for groups on already trained models. For these reasons, we have chosen the following gold standard datasets in the field of \ac{RS}: \texttt{MovieLens1M}~\cite{harper2015movielens}, the most popular dataset in the research of \ac{RS}; \texttt{FilmTrust}~\cite{guo2013novel}, a dataset smaller than \texttt{MovieLens1M} to measure the performance of the aggregation in datasets with a lower number of users, items, and ratings; and \texttt{MyAnimeList}, a dataset with a range of votes higher than the \texttt{MovieLens1M}. Other popular datasets such as \texttt{Netflix Prize} or \texttt{MovieLens10M} have not been selected due to the high computational time required to train and test the models. The main parameters of the selected datasets can be found in \Cref{tab:main_parameters}. 

\begin{table}[ht]
        \centering
        \begin{tabular}{|c|c|c|c|c|c|}
            \hline
            Dataset & \#users & \#items & \#ratings & Scores & Sparsity \\
            \hline
            MovieLens1M & 6,040 & 3,706 & 911,031 & 1 to 5 & 95.94 \\
            \hline
            FilmTrust & 1,508 & 2,071 & 35,497 & 0 to 5 & 87.98 \\
            \hline
            MyAnimeList & 19,179 & 2,692 & 548,967 & 1 to 10 & 98.94 \\
            \hline
        \end{tabular}
        \caption{Main parameters of the datasets used in the experiments.}
        \label{tab:main_parameters}
\end{table}

The generation of synthetic groups has been carried out in such a way that all groups have voted at least 5 items in test. In this way, it is possible to evaluate both the quality of the predictions and the quality of the recommendations to the groups as detailed below. Groups of different sizes (from 2 to 10 users) have been generated. For each group size, 10000 synthetic groups have been generated. The generation of a group has been carried out following the following algorithm:

\begin{enumerate}
    \item Define the size of the group $S$.
    \item Random select 5 items rated in test by at least $S$ users.
    \item Find all users who have rated the 5 items selected in 2.
    \item If we found fewer than $S$ users, go back to 2. Otherwise, random select $S$ users and create a group.
\end{enumerate}

To measure the quality of the predictions for the group, we have calculated the mean absolute error 

\begin{equation}
    MAE = \frac{1}{\#groups} \sum_{G} \frac{1}{\|G\| \cdot \|I_G\|} \sum_{u \in G} \sum_{i \in I_G} \mid \hat{r}_{G,i} - r_{u,i} \mid,
\end{equation}

the mean squared error 

\begin{equation}
    MSE = \frac{1}{\#groups} \sum_{G} \frac{1}{\|G\|} \frac{1}{\|G\| \cdot \|I_G\|} \sum_{u \in G} \sum_{i \in I_G} \left( \hat{r}_{G,i} - r_{u,i} \right)^2, 
\end{equation}

and mean maximum group error

\begin{equation}
    MAX = \frac{1}{\#groups} \sum_{G} \max_{u \in G} \max_{i \in I_G}\mid \hat{r}_{G,i} - r_{u,i} \mid,
\end{equation}

where $I_G$ denotes the items rated by the group $G$

To measure the quality of the recommendations for the group, we have calculated the \ac{NDCG} score

\begin{equation}
    NDCG@N = \frac{1}{\#groups} \sum_{G} \frac{DCG_G@N}{IDCG_G@N},
\label{eq:ndcg}
\end{equation}

\begin{equation}
    DCG_G@N = \sum_{i \in X_G^N} \frac{\bar{r}_{G,i}}{log_2(pos_G(i)+1)},
\label{eq:dcg}
\end{equation}

\begin{equation}
    IDCG_G@N = \sum_{i \in T_G^N} \frac{\bar{r}_{G,i}}{log_2(ipos_G(i)+1)},
\label{eq:idcg}
\end{equation}

where $N$ is the number of items recommended to the group (in our experiments $N=5$ according to the generation of synthetic groups), $X_G^N$ is the set of $N$ items recommended to the group $G$, $pos_G(i)$ is the position of the item $i$ in the group's $G$ recommendation list, $T_G^N$ is the set of the top $N$ items for the group $G$, $ipos_G(i)$ is the ideal rank of the item $i$ for the group $G$, and $\bar{r}_{G,i}$ is the average rating of the users belonging to the group $G$ for the item $i$.

We can see the results of the experiment executed with these scores in \cref{tab:mae} (MAE), \cref{tab:mse} (MSE), \cref{tab:max} (Max), and \cref{tab:ndcg} (NDCG). The cells with the best results have been highlighted, while the standard deviation of each metric is in parentheses. All results are analyzed in~\cref{sec:discussion}.

All experiments have been run using an NVIDIA Quadro RTX 8000 GPU with 48 GB GDDR6 of memory, 4,608 NVIDIA Tensor Cores and a performance of 16.3 TFLOPS. We are committed to reproducible science, so the source code of all experiments with the values of the parameters used and their random seeds have been shared on GitHub\footnote{\url{https://github.com/KNODIS-Research-Group/neural-cf-for-groups}}. 

    \begin{table}[ht]
        \caption{Mean Absolute Error}
        \label{tab:mae}
    
        \begin{subtable}[h]{1\textwidth}
        \resizebox{\textwidth}{!}{%
            \renewcommand{\arraystretch}{1.9}
            \begin{tabular}{|l|c|c|c|c|c|c|c|c|c|}
        \hline
Model \textbackslash Group Size & 2      & 3      & 4      & 5      & 6      & 7      & 8      & 9      & 10     \\ \hline
\makecell[l]{GMF IPA \\ GMF Avg} & \makecell{0.74205\\(0.409)}  & \makecell{0.76075\\(0.341)}  & \makecell{0.76893\\(0.299)}  & \makecell{0.77009\\(0.271)}  & \makecell{0.77745\\(0.249)}  & \makecell{0.77659\\(0.234)}  & \makecell{0.77681\\(0.221)}  & \makecell{0.77599\\(0.212)}  & \makecell{0.77558\\(0.201)} \\ \hline 
GMF Expertise & \makecell{0.74393\\(0.41)}  & \makecell{0.76207\\(0.341)}  & \makecell{0.77018\\(0.299)}  & \makecell{0.77155\\(0.27)}  & \makecell{0.779\\(0.249)}  & \makecell{0.77782\\(0.234)}  & \makecell{0.77834\\(0.221)}  & \makecell{0.77729\\(0.211)}  & \makecell{0.77685\\(0.201)} \\ \hline 
GMF Softmax & \makecell{0.74246\\(0.409)}  & \makecell{0.7608\\(0.341)}  & \makecell{0.76891\\(0.299)}  & \makecell{0.77012\\(0.27)}  & \makecell{0.77751\\(0.249)}  & \makecell{0.7766\\(0.234)}  & \makecell{0.77687\\(0.221)}  & \makecell{0.77602\\(0.212)}  & \makecell{0.77562\\(0.201)} \\ \hline 
MLP IPA & \makecell{0.74956\\(0.444)}  & \makecell{0.77342\\(0.361)}  & \makecell{0.78055\\(0.313)}  & \makecell{0.78211\\(0.28)}  & \makecell{0.78853\\(0.258)}  & \makecell{0.78633\\(0.241)}  & \makecell{0.78678\\(0.228)}  & \makecell{0.78591\\(0.219)}  & \makecell{0.78509\\(0.207)} \\ \hline 
\makecell[l]{MLP Avg \\ DeepGroup} & \cellcolor[gray]{0.8}\makecell{\textbf{0.7236}\\\textbf{(0.486)}}  & \cellcolor[gray]{0.8}\makecell{\textbf{0.74289}\\\textbf{(0.404)}}  & \cellcolor[gray]{0.8}\makecell{\textbf{0.74916}\\\textbf{(0.355)}}  & \cellcolor[gray]{0.8}\makecell{\textbf{0.75018}\\\textbf{(0.32)}}  & \cellcolor[gray]{0.8}\makecell{\textbf{0.75656}\\\textbf{(0.295)}}  & \makecell{0.75537\\(0.275)}  & \cellcolor[gray]{0.8}\makecell{\textbf{0.75551}\\\textbf{(0.261)}}  & \cellcolor[gray]{0.8}\makecell{\textbf{0.75388}\\\textbf{(0.25)}}  & \cellcolor[gray]{0.8}\makecell{\textbf{0.75355}\\\textbf{(0.238)}} \\ \hline 
MLP Expertise & \makecell{0.72596\\(0.486)}  & \makecell{0.74432\\(0.405)}  & \makecell{0.75031\\(0.355)}  & \makecell{0.75132\\(0.32)}  & \makecell{0.75787\\(0.295)}  & \makecell{0.75607\\(0.276)}  & \makecell{0.75709\\(0.261)}  & \makecell{0.75481\\(0.25)}  & \makecell{0.755\\(0.238)} \\ \hline 
MLP Softmax & \makecell{0.72407\\(0.485)}  & \makecell{0.74297\\(0.404)}  & \makecell{0.74925\\(0.355)}  & \makecell{0.75019\\(0.32)}  & \makecell{0.75669\\(0.295)}  & \cellcolor[gray]{0.8}\makecell{\textbf{0.75531}\\\textbf{(0.275)}}  & \makecell{0.7556\\(0.261)}  & \makecell{0.75389\\(0.25)}  & \makecell{0.75361\\(0.238)} \\ \hline 

            \end{tabular}%
}
        \caption{MovieLens1M}
        \label{tab:mae:ml1m}
        \end{subtable}
        \begin{subtable}[h]{1\textwidth}
        \resizebox{\textwidth}{!}{%
            \renewcommand{\arraystretch}{1.9}
            \begin{tabular}{|l|c|c|c|c|c|c|c|c|c|}
        \hline
Model \textbackslash Group Size & 2      & 3      & 4      & 5      & 6      & 7      & 8      & 9      & 10     \\ \hline
\makecell[l]{GMF IPA \\ GMF Avg} & \makecell{0.61552\\(0.451)}  & \makecell{0.71033\\(0.32)}  & \makecell{0.73011\\(0.28)}  & \makecell{0.73419\\(0.252)}  & \makecell{0.73606\\(0.232)}  & \makecell{0.73832\\(0.217)}  & \makecell{0.74045\\(0.203)}  & \makecell{0.74298\\(0.193)}  & \makecell{0.74212\\(0.185)} \\ \hline 
GMF Expertise & \makecell{0.6149\\(0.444)}  & \makecell{0.71239\\(0.319)}  & \makecell{0.73144\\(0.281)}  & \makecell{0.73583\\(0.252)}  & \makecell{0.73742\\(0.232)}  & \makecell{0.7396\\(0.217)}  & \makecell{0.74166\\(0.203)}  & \makecell{0.74418\\(0.193)}  & \makecell{0.74336\\(0.184)} \\ \hline 
GMF Softmax & \makecell{0.61512\\(0.448)}  & \makecell{0.71088\\(0.319)}  & \makecell{0.73035\\(0.28)}  & \makecell{0.73445\\(0.252)}  & \makecell{0.73624\\(0.232)}  & \makecell{0.73847\\(0.217)}  & \makecell{0.74057\\(0.203)}  & \makecell{0.74309\\(0.193)}  & \makecell{0.74222\\(0.185)} \\ \hline 
MLP IPA & \makecell{0.58165\\(0.442)}  & \makecell{0.70368\\(0.325)}  & \makecell{0.72062\\(0.281)}  & \makecell{0.7252\\(0.252)}  & \makecell{0.72788\\(0.232)}  & \makecell{0.73073\\(0.217)}  & \makecell{0.73353\\(0.203)}  & \makecell{0.73623\\(0.193)}  & \makecell{0.73525\\(0.185)} \\ \hline 
\makecell[l]{MLP Avg \\ DeepGroup} & \makecell{0.58199\\(0.447)}  & \cellcolor[gray]{0.8}\makecell{\textbf{0.70129}\\\textbf{(0.319)}}  & \cellcolor[gray]{0.8}\makecell{\textbf{0.71693}\\\textbf{(0.275)}}  & \cellcolor[gray]{0.8}\makecell{\textbf{0.7212}\\\textbf{(0.247)}}  & \cellcolor[gray]{0.8}\makecell{\textbf{0.72421}\\\textbf{(0.228)}}  & \cellcolor[gray]{0.8}\makecell{\textbf{0.727}\\\textbf{(0.214)}}  & \cellcolor[gray]{0.8}\makecell{\textbf{0.72976}\\\textbf{(0.2)}}  & \cellcolor[gray]{0.8}\makecell{\textbf{0.7328}\\\textbf{(0.191)}}  & \cellcolor[gray]{0.8}\makecell{\textbf{0.73188}\\\textbf{(0.183)}} \\ \hline 
MLP Expertise & \cellcolor[gray]{0.8}\makecell{\textbf{0.57903}\\\textbf{(0.453)}}  & \makecell{0.70449\\(0.321)}  & \makecell{0.71891\\(0.276)}  & \makecell{0.72315\\(0.247)}  & \makecell{0.72581\\(0.228)}  & \makecell{0.72865\\(0.213)}  & \makecell{0.73127\\(0.2)}  & \makecell{0.73429\\(0.191)}  & \makecell{0.73336\\(0.183)} \\ \hline 
MLP Softmax & \makecell{0.58063\\(0.45)}  & \makecell{0.70226\\(0.319)}  & \makecell{0.71734\\(0.275)}  & \makecell{0.72153\\(0.247)}  & \makecell{0.72443\\(0.228)}  & \makecell{0.7272\\(0.214)}  & \makecell{0.72993\\(0.2)}  & \makecell{0.73295\\(0.191)}  & \makecell{0.73202\\(0.183)} \\ \hline 

            \end{tabular}%
        }
        \caption{FilmTrust}
        \label{tab:mae:ft}
        \end{subtable}
        \begin{subtable}[h]{1\textwidth}
        \resizebox{\textwidth}{!}{%
            \renewcommand{\arraystretch}{1.9}
            \begin{tabular}{|l|c|c|c|c|c|c|c|c|c|}
        \hline
Model \textbackslash Group Size & 2      & 3      & 4      & 5      & 6      & 7      & 8      & 9      & 10     \\ \hline
\makecell[l]{GMF IPA \\ GMF Avg} & \cellcolor[gray]{0.8}\makecell{\textbf{0.93068}\\\textbf{(0.567)}}  & \makecell{0.95819\\(0.477)}  & \makecell{0.97595\\(0.417)}  & \makecell{0.99149\\(0.38)}  & \makecell{1.0017\\(0.346)}  & \makecell{1.01404\\(0.329)}  & \makecell{1.01942\\(0.312)}  & \makecell{1.022\\(0.297)}  & \makecell{1.02445\\(0.282)} \\ \hline 
GMF Expertise & \makecell{0.93675\\(0.572)}  & \makecell{0.96424\\(0.48)}  & \makecell{0.98092\\(0.419)}  & \makecell{0.99603\\(0.382)}  & \makecell{1.00649\\(0.349)}  & \makecell{1.01805\\(0.331)}  & \makecell{1.02352\\(0.314)}  & \makecell{1.0258\\(0.3)}  & \makecell{1.0284\\(0.284)} \\ \hline 
GMF Softmax & \makecell{0.93219\\(0.568)}  & \makecell{0.95848\\(0.477)}  & \makecell{0.97559\\(0.417)}  & \makecell{0.99104\\(0.38)}  & \makecell{1.00133\\(0.346)}  & \makecell{1.01363\\(0.329)}  & \makecell{1.0191\\(0.312)}  & \makecell{1.02173\\(0.297)}  & \makecell{1.02422\\(0.282)} \\ \hline 
MLP IPA & \makecell{0.95479\\(0.585)}  & \makecell{0.98155\\(0.484)}  & \makecell{0.99709\\(0.42)}  & \makecell{1.00977\\(0.381)}  & \makecell{1.01745\\(0.347)}  & \makecell{1.02794\\(0.329)}  & \makecell{1.03188\\(0.311)}  & \makecell{1.03335\\(0.298)}  & \makecell{1.03451\\(0.282)} \\ \hline 
\makecell[l]{MLP Avg \\ DeepGroup} & \makecell{0.93161\\(0.618)}  & \makecell{0.95609\\(0.515)}  & \cellcolor[gray]{0.8}\makecell{\textbf{0.96961}\\\textbf{(0.45)}}  & \makecell{0.98456\\(0.408)}  & \makecell{0.99331\\(0.371)}  & \makecell{1.0052\\(0.351)}  & \makecell{1.00857\\(0.332)}  & \makecell{1.01039\\(0.317)}  & \makecell{1.01233\\(0.3)} \\ \hline 
MLP Expertise & \makecell{0.93803\\(0.622)}  & \makecell{0.96132\\(0.517)}  & \makecell{0.97587\\(0.453)}  & \makecell{0.98923\\(0.409)}  & \makecell{0.99851\\(0.373)}  & \makecell{1.00879\\(0.353)}  & \makecell{1.01258\\(0.334)}  & \makecell{1.01493\\(0.319)}  & \makecell{1.01599\\(0.302)} \\ \hline 
MLP Softmax & \makecell{0.93351\\(0.619)}  & \cellcolor[gray]{0.8}\makecell{\textbf{0.95594}\\\textbf{(0.515)}}  & \makecell{0.97007\\(0.451)}  & \cellcolor[gray]{0.8}\makecell{\textbf{0.9843}\\\textbf{(0.408)}}  & \cellcolor[gray]{0.8}\makecell{\textbf{0.99329}\\\textbf{(0.371)}}  & \cellcolor[gray]{0.8}\makecell{\textbf{1.00486}\\\textbf{(0.351)}}  & \cellcolor[gray]{0.8}\makecell{\textbf{1.00844}\\\textbf{(0.332)}}  & \cellcolor[gray]{0.8}\makecell{\textbf{1.0101}\\\textbf{(0.317)}}  & \cellcolor[gray]{0.8}\makecell{\textbf{1.01211}\\\textbf{(0.3)}} \\ \hline 

            \end{tabular}%
        }
        \caption{MyAnimeList}
        \label{tab:mae:anime}
        \end{subtable}
    \end{table}

    \begin{table}[ht]
        \caption{Mean Squared Error}
        \label{tab:mse}
    
        \begin{subtable}[h]{1\textwidth}
        \resizebox{\textwidth}{!}{%
            \renewcommand{\arraystretch}{1.9}
            \begin{tabular}{|l|c|c|c|c|c|c|c|c|c|}
        \hline
Model \textbackslash Group Size & 2      & 3      & 4      & 5      & 6      & 7      & 8      & 9      & 10     \\ \hline
\makecell[l]{GMF IPA \\ GMF Avg} & \cellcolor[gray]{0.8}\makecell{\textbf{0.87504}\\\textbf{(0.914)}}  & \cellcolor[gray]{0.8}\makecell{\textbf{0.90947}\\\textbf{(0.767)}}  & \makecell{0.92437\\(0.676)}  & \makecell{0.92648\\(0.609)}  & \makecell{0.94191\\(0.567)}  & \makecell{0.93896\\(0.53)}  & \makecell{0.9376\\(0.5)}  & \makecell{0.93832\\(0.48)}  & \makecell{0.93571\\(0.454)} \\ \hline 
GMF Expertise & \makecell{0.88044\\(0.918)}  & \makecell{0.91329\\(0.77)}  & \makecell{0.92754\\(0.678)}  & \makecell{0.92979\\(0.61)}  & \makecell{0.94504\\(0.568)}  & \makecell{0.94135\\(0.531)}  & \makecell{0.93994\\(0.501)}  & \makecell{0.94037\\(0.481)}  & \makecell{0.93737\\(0.454)} \\ \hline 
GMF Softmax & \makecell{0.87614\\(0.914)}  & \makecell{0.90952\\(0.767)}  & \cellcolor[gray]{0.8}\makecell{\textbf{0.92424}\\\textbf{(0.676)}}  & \cellcolor[gray]{0.8}\makecell{\textbf{0.92645}\\\textbf{(0.609)}}  & \cellcolor[gray]{0.8}\makecell{\textbf{0.94188}\\\textbf{(0.567)}}  & \cellcolor[gray]{0.8}\makecell{\textbf{0.93885}\\\textbf{(0.53)}}  & \cellcolor[gray]{0.8}\makecell{\textbf{0.93752}\\\textbf{(0.5)}}  & \cellcolor[gray]{0.8}\makecell{\textbf{0.93823}\\\textbf{(0.48)}}  & \cellcolor[gray]{0.8}\makecell{\textbf{0.93561}\\\textbf{(0.454)}} \\ \hline 
MLP IPA & \makecell{0.94301\\(0.982)}  & \makecell{0.96557\\(0.814)}  & \makecell{0.96809\\(0.712)}  & \makecell{0.96499\\(0.635)}  & \makecell{0.97727\\(0.591)}  & \makecell{0.96889\\(0.55)}  & \makecell{0.96634\\(0.518)}  & \makecell{0.96648\\(0.498)}  & \makecell{0.96187\\(0.47)} \\ \hline 
\makecell[l]{MLP Avg \\ DeepGroup} & \makecell{0.99415\\(1.037)}  & \makecell{1.03182\\(0.875)}  & \makecell{1.04278\\(0.779)}  & \makecell{1.04242\\(0.695)}  & \makecell{1.05597\\(0.651)}  & \makecell{1.05056\\(0.605)}  & \makecell{1.04927\\(0.574)}  & \makecell{1.04836\\(0.552)}  & \makecell{1.04669\\(0.524)} \\ \hline 
MLP Expertise & \makecell{0.9986\\(1.038)}  & \makecell{1.03522\\(0.88)}  & \makecell{1.04511\\(0.78)}  & \makecell{1.04435\\(0.696)}  & \makecell{1.05806\\(0.651)}  & \makecell{1.05121\\(0.604)}  & \makecell{1.05084\\(0.575)}  & \makecell{1.0491\\(0.553)}  & \makecell{1.04787\\(0.523)} \\ \hline 
MLP Softmax & \makecell{0.99487\\(1.035)}  & \makecell{1.03145\\(0.874)}  & \makecell{1.04258\\(0.779)}  & \makecell{1.04235\\(0.694)}  & \makecell{1.05605\\(0.651)}  & \makecell{1.05034\\(0.604)}  & \makecell{1.04925\\(0.574)}  & \makecell{1.04822\\(0.552)}  & \makecell{1.04659\\(0.524)} \\ \hline 

            \end{tabular}%
        }
        \caption{MovieLens1M}
        \label{tab:mse:ml1m}
        \end{subtable}
        \begin{subtable}[h]{1\textwidth}
        \resizebox{\textwidth}{!}{%
            \renewcommand{\arraystretch}{1.9}
            \begin{tabular}{|l|c|c|c|c|c|c|c|c|c|}
        \hline
Model \textbackslash Group Size & 2      & 3      & 4      & 5      & 6      & 7      & 8      & 9      & 10     \\ \hline
\makecell[l]{GMF IPA \\ GMF Avg} & \makecell{0.67941\\(1.077)}  & \makecell{0.7889\\(0.688)}  & \makecell{0.82014\\(0.608)}  & \makecell{0.82302\\(0.544)}  & \makecell{0.82574\\(0.502)}  & \makecell{0.83038\\(0.469)}  & \makecell{0.8324\\(0.44)}  & \makecell{0.8375\\(0.42)}  & \makecell{0.83544\\(0.401)} \\ \hline 
GMF Expertise & \makecell{0.67314\\(1.05)}  & \makecell{0.79105\\(0.686)}  & \makecell{0.82229\\(0.608)}  & \makecell{0.82546\\(0.543)}  & \makecell{0.82758\\(0.501)}  & \makecell{0.83205\\(0.468)}  & \makecell{0.83382\\(0.439)}  & \makecell{0.83875\\(0.418)}  & \makecell{0.83673\\(0.399)} \\ \hline 
GMF Softmax & \makecell{0.67596\\(1.063)}  & \makecell{0.7892\\(0.687)}  & \makecell{0.82043\\(0.608)}  & \makecell{0.82333\\(0.544)}  & \makecell{0.82592\\(0.502)}  & \makecell{0.83053\\(0.469)}  & \makecell{0.83251\\(0.44)}  & \makecell{0.83758\\(0.42)}  & \makecell{0.83552\\(0.401)} \\ \hline 
MLP IPA & \cellcolor[gray]{0.8}\makecell{\textbf{0.61169}\\\textbf{(1.002)}}  & \makecell{0.77197\\(0.683)}  & \makecell{0.79514\\(0.592)}  & \makecell{0.7988\\(0.528)}  & \makecell{0.80315\\(0.487)}  & \makecell{0.80807\\(0.454)}  & \makecell{0.81084\\(0.427)}  & \makecell{0.81611\\(0.406)}  & \makecell{0.81408\\(0.388)} \\ \hline 
\makecell[l]{MLP Avg \\ DeepGroup} & \makecell{0.61859\\(1.009)}  & \cellcolor[gray]{0.8}\makecell{\textbf{0.7636}\\\textbf{(0.674)}}  & \cellcolor[gray]{0.8}\makecell{\textbf{0.78665}\\\textbf{(0.585)}}  & \cellcolor[gray]{0.8}\makecell{\textbf{0.79085}\\\textbf{(0.523)}}  & \cellcolor[gray]{0.8}\makecell{\textbf{0.79606}\\\textbf{(0.484)}}  & \cellcolor[gray]{0.8}\makecell{\textbf{0.80126}\\\textbf{(0.452)}}  & \cellcolor[gray]{0.8}\makecell{\textbf{0.80444}\\\textbf{(0.425)}}  & \cellcolor[gray]{0.8}\makecell{\textbf{0.81023}\\\textbf{(0.405)}}  & \cellcolor[gray]{0.8}\makecell{\textbf{0.80841}\\\textbf{(0.387)}} \\ \hline 
MLP Expertise & \makecell{0.62484\\(0.979)}  & \makecell{0.7691\\(0.68)}  & \makecell{0.78956\\(0.586)}  & \makecell{0.79332\\(0.522)}  & \makecell{0.79799\\(0.483)}  & \makecell{0.80305\\(0.45)}  & \makecell{0.80595\\(0.423)}  & \makecell{0.81166\\(0.403)}  & \makecell{0.80977\\(0.385)} \\ \hline 
MLP Softmax & \makecell{0.62106\\(0.992)}  & \makecell{0.7649\\(0.675)}  & \makecell{0.78709\\(0.585)}  & \makecell{0.79116\\(0.523)}  & \makecell{0.79625\\(0.484)}  & \makecell{0.80142\\(0.452)}  & \makecell{0.80456\\(0.424)}  & \makecell{0.81033\\(0.404)}  & \makecell{0.8085\\(0.387)} \\ \hline 

            \end{tabular}%
        }
        \caption{FilmTrust}
        \label{tab:mse:ft}
        \end{subtable}
        \begin{subtable}[h]{1\textwidth}
        \resizebox{\textwidth}{!}{%
            \renewcommand{\arraystretch}{1.9}
            \begin{tabular}{|l|c|c|c|c|c|c|c|c|c|}
        \hline
Model \textbackslash Group Size & 2      & 3      & 4      & 5      & 6      & 7      & 8      & 9      & 10     \\ \hline
\makecell[l]{GMF IPA \\ GMF Avg} & \cellcolor[gray]{0.8}\makecell{\textbf{1.47037}\\\textbf{(1.922)}}  & \cellcolor[gray]{0.8}\makecell{\textbf{1.53981}\\\textbf{(1.648)}}  & \cellcolor[gray]{0.8}\makecell{\textbf{1.58373}\\\textbf{(1.43)}}  & \cellcolor[gray]{0.8}\makecell{\textbf{1.63441}\\\textbf{(1.324)}}  & \cellcolor[gray]{0.8}\makecell{\textbf{1.66233}\\\textbf{(1.213)}}  & \cellcolor[gray]{0.8}\makecell{\textbf{1.70615}\\\textbf{(1.178)}}  & \cellcolor[gray]{0.8}\makecell{\textbf{1.72188}\\\textbf{(1.111)}}  & \cellcolor[gray]{0.8}\makecell{\textbf{1.73177}\\\textbf{(1.065)}}  & \cellcolor[gray]{0.8}\makecell{\textbf{1.74048}\\\textbf{(1.015)}} \\ \hline 
GMF Expertise & \makecell{1.49874\\(1.963)}  & \makecell{1.57277\\(1.689)}  & \makecell{1.61625\\(1.463)}  & \makecell{1.66444\\(1.351)}  & \makecell{1.69199\\(1.239)}  & \makecell{1.73293\\(1.202)}  & \makecell{1.74812\\(1.133)}  & \makecell{1.75649\\(1.087)}  & \makecell{1.76487\\(1.035)} \\ \hline 
GMF Softmax & \makecell{1.47707\\(1.932)}  & \makecell{1.54383\\(1.654)}  & \makecell{1.58617\\(1.433)}  & \makecell{1.63575\\(1.326)}  & \makecell{1.66342\\(1.214)}  & \makecell{1.70677\\(1.179)}  & \makecell{1.72239\\(1.112)}  & \makecell{1.73223\\(1.066)}  & \makecell{1.74092\\(1.016)} \\ \hline 
MLP IPA & \makecell{1.56623\\(1.959)}  & \makecell{1.61737\\(1.655)}  & \makecell{1.64909\\(1.431)}  & \makecell{1.69132\\(1.32)}  & \makecell{1.71029\\(1.209)}  & \makecell{1.74683\\(1.169)}  & \makecell{1.75711\\(1.105)}  & \makecell{1.76368\\(1.058)}  & \makecell{1.76877\\(1.008)} \\ \hline 
\makecell[l]{MLP Avg \\ DeepGroup} & \makecell{1.61669\\(2.032)}  & \makecell{1.68103\\(1.718)}  & \makecell{1.71103\\(1.492)}  & \makecell{1.75681\\(1.368)}  & \makecell{1.77828\\(1.254)}  & \makecell{1.81759\\(1.217)}  & \makecell{1.82764\\(1.148)}  & \makecell{1.83403\\(1.098)}  & \makecell{1.84177\\(1.049)} \\ \hline 
MLP Expertise & \makecell{1.64412\\(2.059)}  & \makecell{1.70965\\(1.747)}  & \makecell{1.74467\\(1.517)}  & \makecell{1.78446\\(1.386)}  & \makecell{1.80671\\(1.276)}  & \makecell{1.84221\\(1.232)}  & \makecell{1.85095\\(1.163)}  & \makecell{1.85803\\(1.116)}  & \makecell{1.86351\\(1.063)} \\ \hline 
MLP Softmax & \makecell{1.62458\\(2.038)}  & \makecell{1.68445\\(1.723)}  & \makecell{1.7153\\(1.496)}  & \makecell{1.75838\\(1.368)}  & \makecell{1.78024\\(1.256)}  & \makecell{1.81848\\(1.218)}  & \makecell{1.82855\\(1.149)}  & \makecell{1.83446\\(1.099)}  & \makecell{1.84249\\(1.049)} \\ \hline 

            \end{tabular}%
        }
        \caption{MyAnimeList}
        \label{tab:mse:anime}
        \end{subtable}
    \end{table}

    \begin{table}[ht]
        \caption{Mean Max Error}
        \label{tab:max}
    
        \begin{subtable}[h]{1\textwidth}
        \resizebox{\textwidth}{!}{%
            \renewcommand{\arraystretch}{1.9}
            \begin{tabular}{|l|c|c|c|c|c|c|c|c|c|}
        \hline
Model \textbackslash Group Size & 2      & 3      & 4      & 5      & 6      & 7      & 8      & 9      & 10     \\ \hline
\makecell[l]{GMF IPA \\ GMF Avg} & \cellcolor[gray]{0.8}\makecell{\textbf{1.02112}\\\textbf{(0.6)}}  & \cellcolor[gray]{0.8}\makecell{\textbf{1.2213}\\\textbf{(0.598)}}  & \makecell{1.35658\\(0.588)}  & \cellcolor[gray]{0.8}\makecell{\textbf{1.45195}\\\textbf{(0.583)}}  & \cellcolor[gray]{0.8}\makecell{\textbf{1.54115}\\\textbf{(0.578)}}  & \makecell{1.59908\\(0.577)}  & \makecell{1.64979\\(0.573)}  & \makecell{1.69807\\(0.576)}  & \makecell{1.73598\\(0.571)} \\ \hline 
GMF Expertise & \makecell{1.02474\\(0.602)}  & \makecell{1.22441\\(0.599)}  & \makecell{1.35928\\(0.59)}  & \makecell{1.4551\\(0.584)}  & \makecell{1.54414\\(0.579)}  & \makecell{1.60112\\(0.577)}  & \makecell{1.65118\\(0.574)}  & \makecell{1.69926\\(0.576)}  & \makecell{1.7366\\(0.571)} \\ \hline 
GMF Softmax & \makecell{1.02191\\(0.6)}  & \makecell{1.22143\\(0.598)}  & \cellcolor[gray]{0.8}\makecell{\textbf{1.35653}\\\textbf{(0.588)}}  & \makecell{1.45202\\(0.583)}  & \makecell{1.54118\\(0.578)}  & \cellcolor[gray]{0.8}\makecell{\textbf{1.59897}\\\textbf{(0.577)}}  & \cellcolor[gray]{0.8}\makecell{\textbf{1.64965}\\\textbf{(0.573)}}  & \cellcolor[gray]{0.8}\makecell{\textbf{1.69792}\\\textbf{(0.576)}}  & \cellcolor[gray]{0.8}\makecell{\textbf{1.7358}\\\textbf{(0.571)}} \\ \hline 
MLP IPA & \makecell{1.04098\\(0.642)}  & \makecell{1.25715\\(0.614)}  & \makecell{1.38214\\(0.602)}  & \makecell{1.47972\\(0.591)}  & \makecell{1.56516\\(0.586)}  & \makecell{1.62067\\(0.581)}  & \makecell{1.66905\\(0.576)}  & \makecell{1.71687\\(0.58)}  & \makecell{1.7528\\(0.573)} \\ \hline 
\makecell[l]{MLP Avg \\ DeepGroup} & \makecell{1.07144\\(0.666)}  & \makecell{1.28743\\(0.643)}  & \makecell{1.41937\\(0.635)}  & \makecell{1.51234\\(0.633)}  & \makecell{1.59812\\(0.638)}  & \makecell{1.65421\\(0.642)}  & \makecell{1.70673\\(0.641)}  & \makecell{1.75552\\(0.647)}  & \makecell{1.79703\\(0.645)} \\ \hline 
MLP Expertise & \makecell{1.07476\\(0.667)}  & \makecell{1.28932\\(0.644)}  & \makecell{1.42078\\(0.635)}  & \makecell{1.51325\\(0.634)}  & \makecell{1.59883\\(0.638)}  & \makecell{1.65371\\(0.64)}  & \makecell{1.70637\\(0.64)}  & \makecell{1.75428\\(0.645)}  & \makecell{1.79566\\(0.643)} \\ \hline 
MLP Softmax & \makecell{1.07226\\(0.666)}  & \makecell{1.28716\\(0.643)}  & \makecell{1.41906\\(0.635)}  & \makecell{1.51227\\(0.633)}  & \makecell{1.59788\\(0.638)}  & \makecell{1.65391\\(0.641)}  & \makecell{1.7066\\(0.641)}  & \makecell{1.75515\\(0.646)}  & \makecell{1.79669\\(0.645)} \\ \hline 

            \end{tabular}%
        }
        \caption{MovieLens1M}
        \label{tab:max:ml1m}
        \end{subtable}
        \begin{subtable}[h]{1\textwidth}
        \resizebox{\textwidth}{!}{%
            \renewcommand{\arraystretch}{1.9}
            \begin{tabular}{|l|c|c|c|c|c|c|c|c|c|}
        \hline
Model \textbackslash Group Size & 2      & 3      & 4      & 5      & 6      & 7      & 8      & 9      & 10     \\ \hline
\makecell[l]{GMF IPA \\ GMF Avg} & \makecell{0.85192\\(0.555)}  & \makecell{1.13757\\(0.573)}  & \makecell{1.27696\\(0.573)}  & \makecell{1.36783\\(0.57)}  & \makecell{1.44244\\(0.571)}  & \makecell{1.51262\\(0.575)}  & \makecell{1.56818\\(0.573)}  & \makecell{1.62204\\(0.571)}  & \makecell{1.6636\\(0.567)} \\ \hline 
GMF Expertise & \makecell{0.85171\\(0.549)}  & \makecell{1.13845\\(0.571)}  & \makecell{1.27815\\(0.571)}  & \makecell{1.36895\\(0.567)}  & \makecell{1.4425\\(0.569)}  & \makecell{1.51199\\(0.573)}  & \makecell{1.56717\\(0.57)}  & \makecell{1.62058\\(0.569)}  & \makecell{1.66146\\(0.565)} \\ \hline 
GMF Softmax & \makecell{0.85147\\(0.552)}  & \makecell{1.13761\\(0.572)}  & \makecell{1.27708\\(0.572)}  & \makecell{1.3679\\(0.569)}  & \makecell{1.44234\\(0.571)}  & \makecell{1.51245\\(0.575)}  & \makecell{1.56799\\(0.572)}  & \makecell{1.62184\\(0.571)}  & \makecell{1.66335\\(0.567)} \\ \hline 
MLP IPA & \cellcolor[gray]{0.8}\makecell{\textbf{0.78405}\\\textbf{(0.555)}}  & \makecell{1.11519\\(0.564)}  & \makecell{1.24935\\(0.56)}  & \makecell{1.33899\\(0.555)}  & \makecell{1.41348\\(0.556)}  & \makecell{1.48001\\(0.557)}  & \makecell{1.5341\\(0.554)}  & \makecell{1.58622\\(0.552)}  & \makecell{1.62791\\(0.549)} \\ \hline 
\makecell[l]{MLP Avg \\ DeepGroup} & \makecell{0.79205\\(0.553)}  & \cellcolor[gray]{0.8}\makecell{\textbf{1.11222}\\\textbf{(0.557)}}  & \cellcolor[gray]{0.8}\makecell{\textbf{1.24695}\\\textbf{(0.557)}}  & \makecell{1.33662\\(0.555)}  & \makecell{1.4118\\(0.557)}  & \makecell{1.47902\\(0.559)}  & \makecell{1.53438\\(0.556)}  & \makecell{1.58697\\(0.555)}  & \makecell{1.62916\\(0.55)} \\ \hline 
MLP Expertise & \makecell{0.79191\\(0.56)}  & \makecell{1.11441\\(0.556)}  & \makecell{1.24814\\(0.555)}  & \makecell{1.33715\\(0.551)}  & \cellcolor[gray]{0.8}\makecell{\textbf{1.41142}\\\textbf{(0.554)}}  & \cellcolor[gray]{0.8}\makecell{\textbf{1.47765}\\\textbf{(0.556)}}  & \cellcolor[gray]{0.8}\makecell{\textbf{1.53235}\\\textbf{(0.553)}}  & \cellcolor[gray]{0.8}\makecell{\textbf{1.58432}\\\textbf{(0.551)}}  & \cellcolor[gray]{0.8}\makecell{\textbf{1.62581}\\\textbf{(0.547)}} \\ \hline 
MLP Softmax & \makecell{0.79235\\(0.555)}  & \makecell{1.11261\\(0.556)}  & \makecell{1.24698\\(0.556)}  & \cellcolor[gray]{0.8}\makecell{\textbf{1.33654}\\\textbf{(0.554)}}  & \makecell{1.41161\\(0.556)}  & \makecell{1.47872\\(0.558)}  & \makecell{1.53406\\(0.555)}  & \makecell{1.58662\\(0.554)}  & \makecell{1.62879\\(0.55)} \\ \hline 

            \end{tabular}%
        }
        \caption{FilmTrust}
        \label{tab:max:ft}
        \end{subtable}
        \begin{subtable}[h]{1\textwidth}
        \resizebox{\textwidth}{!}{%
            \renewcommand{\arraystretch}{1.9}
            \begin{tabular}{|l|c|c|c|c|c|c|c|c|c|}
        \hline
Model \textbackslash Group Size & 2      & 3      & 4      & 5      & 6      & 7      & 8      & 9      & 10     \\ \hline
\makecell[l]{GMF IPA \\ GMF Avg} & \cellcolor[gray]{0.8}\makecell{\textbf{1.30264}\\\textbf{(0.845)}}  & \cellcolor[gray]{0.8}\makecell{\textbf{1.58715}\\\textbf{(0.866)}}  & \cellcolor[gray]{0.8}\makecell{\textbf{1.79755}\\\textbf{(0.864)}}  & \cellcolor[gray]{0.8}\makecell{\textbf{1.9702}\\\textbf{(0.877)}}  & \cellcolor[gray]{0.8}\makecell{\textbf{2.10648}\\\textbf{(0.876)}}  & \cellcolor[gray]{0.8}\makecell{\textbf{2.22836}\\\textbf{(0.896)}}  & \cellcolor[gray]{0.8}\makecell{\textbf{2.31758}\\\textbf{(0.895)}}  & \cellcolor[gray]{0.8}\makecell{\textbf{2.39541}\\\textbf{(0.91)}}  & \cellcolor[gray]{0.8}\makecell{\textbf{2.47078}\\\textbf{(0.919)}} \\ \hline 
GMF Expertise & \makecell{1.31784\\(0.858)}  & \makecell{1.60964\\(0.883)}  & \makecell{1.8266\\(0.882)}  & \makecell{1.99902\\(0.894)}  & \makecell{2.13631\\(0.891)}  & \makecell{2.25748\\(0.91)}  & \makecell{2.34706\\(0.908)}  & \makecell{2.42357\\(0.922)}  & \makecell{2.49875\\(0.932)} \\ \hline 
GMF Softmax & \makecell{1.30626\\(0.847)}  & \makecell{1.59061\\(0.869)}  & \makecell{1.80139\\(0.867)}  & \makecell{1.97309\\(0.879)}  & \makecell{2.10896\\(0.878)}  & \makecell{2.23046\\(0.898)}  & \makecell{2.31946\\(0.896)}  & \makecell{2.39701\\(0.911)}  & \makecell{2.47222\\(0.92)} \\ \hline 
MLP IPA & \makecell{1.34151\\(0.874)}  & \makecell{1.63805\\(0.873)}  & \makecell{1.84317\\(0.862)}  & \makecell{2.01188\\(0.868)}  & \makecell{2.1406\\(0.863)}  & \makecell{2.25649\\(0.876)}  & \makecell{2.33763\\(0.875)}  & \makecell{2.4112\\(0.889)}  & \makecell{2.4836\\(0.896)} \\ \hline 
\makecell[l]{MLP Avg \\ DeepGroup} & \makecell{1.36617\\(0.893)}  & \makecell{1.66385\\(0.902)}  & \makecell{1.8675\\(0.896)}  & \makecell{2.03818\\(0.907)}  & \makecell{2.17095\\(0.906)}  & \makecell{2.28936\\(0.924)}  & \makecell{2.37588\\(0.921)}  & \makecell{2.4495\\(0.934)}  & \makecell{2.52558\\(0.943)} \\ \hline 
MLP Expertise & \makecell{1.38038\\(0.9)}  & \makecell{1.68165\\(0.913)}  & \makecell{1.89237\\(0.908)}  & \makecell{2.0611\\(0.918)}  & \makecell{2.1951\\(0.914)}  & \makecell{2.31311\\(0.929)}  & \makecell{2.39796\\(0.923)}  & \makecell{2.47081\\(0.935)}  & \makecell{2.54585\\(0.943)} \\ \hline 
MLP Softmax & \makecell{1.37076\\(0.894)}  & \makecell{1.66676\\(0.904)}  & \makecell{1.87147\\(0.898)}  & \makecell{2.04076\\(0.908)}  & \makecell{2.17377\\(0.907)}  & \makecell{2.29111\\(0.924)}  & \makecell{2.37734\\(0.921)}  & \makecell{2.45073\\(0.934)}  & \makecell{2.52696\\(0.943)} \\ \hline 

            \end{tabular}%
        }
        \caption{MyAnimeList}
        \label{tab:max:anime}
        \end{subtable}
    \end{table}

    \begin{table}[ht]
        \caption{Discounted cumulative gain}
        \label{tab:ndcg}
    
        \begin{subtable}[h]{1\textwidth}
        \resizebox{\textwidth}{!}{%
            \renewcommand{\arraystretch}{1.9}
            \begin{tabular}{|l|c|c|c|c|c|c|c|c|c|}
        \hline
Model \textbackslash Group Size & 2      & 3      & 4      & 5      & 6      & 7      & 8      & 9      & 10     \\ \hline
\makecell[l]{GMF IPA \\ GMF Avg} & \cellcolor[gray]{0.8}\makecell{\textbf{0.98021}\\\textbf{(0.024)}}  & \cellcolor[gray]{0.8}\makecell{\textbf{0.98543}\\\textbf{(0.018)}}  & \makecell{0.9886\\(0.015)}  & \cellcolor[gray]{0.8}\makecell{\textbf{0.99067}\\\textbf{(0.012)}}  & \makecell{0.99178\\(0.011)}  & \makecell{0.99286\\(0.01)}  & \makecell{0.99358\\(0.009)}  & \cellcolor[gray]{0.8}\makecell{\textbf{0.99415}\\\textbf{(0.008)}}  & \cellcolor[gray]{0.8}\makecell{\textbf{0.99457}\\\textbf{(0.008)}} \\ \hline 
GMF Expertise & \makecell{0.97995\\(0.024)}  & \makecell{0.98528\\(0.019)}  & \makecell{0.98853\\(0.015)}  & \makecell{0.99058\\(0.012)}  & \makecell{0.99171\\(0.011)}  & \makecell{0.99282\\(0.01)}  & \makecell{0.99351\\(0.009)}  & \makecell{0.99411\\(0.008)}  & \makecell{0.99454\\(0.008)} \\ \hline 
GMF Softmax & \makecell{0.98006\\(0.024)}  & \makecell{0.98536\\(0.019)}  & \cellcolor[gray]{0.8}\makecell{\textbf{0.9886}\\\textbf{(0.015)}}  & \makecell{0.99064\\(0.012)}  & \cellcolor[gray]{0.8}\makecell{\textbf{0.99178}\\\textbf{(0.011)}}  & \cellcolor[gray]{0.8}\makecell{\textbf{0.99288}\\\textbf{(0.01)}}  & \cellcolor[gray]{0.8}\makecell{\textbf{0.9936}\\\textbf{(0.009)}}  & \makecell{0.99415\\(0.008)}  & \makecell{0.99456\\(0.008)} \\ \hline 
MLP IPA & \makecell{0.97854\\(0.026)}  & \makecell{0.98342\\(0.02)}  & \makecell{0.98689\\(0.016)}  & \makecell{0.98906\\(0.014)}  & \makecell{0.99046\\(0.012)}  & \makecell{0.99178\\(0.011)}  & \makecell{0.99251\\(0.01)}  & \makecell{0.99303\\(0.009)}  & \makecell{0.99358\\(0.009)} \\ \hline 
\makecell[l]{MLP Avg \\ DeepGroup} & \makecell{0.97778\\(0.026)}  & \makecell{0.98247\\(0.021)}  & \makecell{0.98556\\(0.017)}  & \makecell{0.98762\\(0.015)}  & \makecell{0.98894\\(0.014)}  & \makecell{0.98999\\(0.013)}  & \makecell{0.99064\\(0.012)}  & \makecell{0.99109\\(0.011)}  & \makecell{0.99152\\(0.011)} \\ \hline 
MLP Expertise & \makecell{0.97777\\(0.026)}  & \makecell{0.98251\\(0.021)}  & \makecell{0.9855\\(0.017)}  & \makecell{0.98763\\(0.015)}  & \makecell{0.98894\\(0.014)}  & \makecell{0.99004\\(0.013)}  & \makecell{0.99071\\(0.012)}  & \makecell{0.99116\\(0.011)}  & \makecell{0.9916\\(0.011)} \\ \hline 
MLP Softmax & \makecell{0.97784\\(0.026)}  & \makecell{0.98248\\(0.021)}  & \makecell{0.98554\\(0.017)}  & \makecell{0.98762\\(0.015)}  & \makecell{0.98895\\(0.014)}  & \makecell{0.98998\\(0.013)}  & \makecell{0.99069\\(0.012)}  & \makecell{0.9911\\(0.011)}  & \makecell{0.99153\\(0.011)} \\ \hline 

            \end{tabular}%
        }
        \caption{MovieLens1M}
        \label{tab:ndcg:ml1m}
        \end{subtable}
        \begin{subtable}[h]{1\textwidth}
        \resizebox{\textwidth}{!}{%
            \renewcommand{\arraystretch}{1.9}
            \begin{tabular}{|l|c|c|c|c|c|c|c|c|c|}
        \hline
Model \textbackslash Group Size & 2      & 3      & 4      & 5      & 6      & 7      & 8      & 9      & 10     \\ \hline
\makecell[l]{GMF IPA \\ GMF Avg} & \makecell{0.96788\\(0.028)}  & \makecell{0.97795\\(0.022)}  & \makecell{0.9815\\(0.019)}  & \makecell{0.98393\\(0.016)}  & \makecell{0.98596\\(0.014)}  & \makecell{0.9876\\(0.013)}  & \makecell{0.98869\\(0.011)}  & \makecell{0.98967\\(0.011)}  & \makecell{0.99038\\(0.01)} \\ \hline 
GMF Expertise & \makecell{0.96795\\(0.028)}  & \makecell{0.97794\\(0.022)}  & \makecell{0.9815\\(0.019)}  & \makecell{0.98392\\(0.016)}  & \makecell{0.98593\\(0.014)}  & \makecell{0.9876\\(0.013)}  & \makecell{0.98868\\(0.012)}  & \makecell{0.98965\\(0.011)}  & \makecell{0.99038\\(0.01)} \\ \hline 
GMF Softmax & \makecell{0.96795\\(0.028)}  & \makecell{0.97796\\(0.022)}  & \makecell{0.9815\\(0.019)}  & \makecell{0.98394\\(0.016)}  & \makecell{0.98596\\(0.014)}  & \makecell{0.9876\\(0.013)}  & \makecell{0.98868\\(0.012)}  & \makecell{0.98967\\(0.011)}  & \makecell{0.99038\\(0.01)} \\ \hline 
MLP IPA & \cellcolor[gray]{0.8}\makecell{\textbf{0.97057}\\\textbf{(0.027)}}  & \makecell{0.9788\\(0.023)}  & \makecell{0.98249\\(0.019)}  & \makecell{0.98523\\(0.016)}  & \makecell{0.98704\\(0.015)}  & \cellcolor[gray]{0.8}\makecell{\textbf{0.98896}\\\textbf{(0.012)}}  & \makecell{0.98994\\(0.011)}  & \cellcolor[gray]{0.8}\makecell{\textbf{0.9911}\\\textbf{(0.01)}}  & \makecell{0.99165\\(0.01)} \\ \hline 
\makecell[l]{MLP Avg \\ DeepGroup} & \makecell{0.96897\\(0.027)}  & \makecell{0.9789\\(0.022)}  & \makecell{0.98275\\(0.018)}  & \makecell{0.98527\\(0.016)}  & \cellcolor[gray]{0.8}\makecell{\textbf{0.98729}\\\textbf{(0.014)}}  & \makecell{0.98895\\(0.012)}  & \cellcolor[gray]{0.8}\makecell{\textbf{0.98998}\\\textbf{(0.011)}}  & \makecell{0.99104\\(0.01)}  & \makecell{0.99166\\(0.009)} \\ \hline 
MLP Expertise & \makecell{0.9694\\(0.027)}  & \cellcolor[gray]{0.8}\makecell{\textbf{0.97897}\\\textbf{(0.022)}}  & \makecell{0.98276\\(0.018)}  & \cellcolor[gray]{0.8}\makecell{\textbf{0.98532}\\\textbf{(0.016)}}  & \makecell{0.98724\\(0.014)}  & \makecell{0.98893\\(0.012)}  & \makecell{0.98993\\(0.011)}  & \makecell{0.99106\\(0.01)}  & \makecell{0.99166\\(0.009)} \\ \hline 
MLP Softmax & \makecell{0.9693\\(0.027)}  & \makecell{0.97891\\(0.022)}  & \cellcolor[gray]{0.8}\makecell{\textbf{0.98278}\\\textbf{(0.018)}}  & \makecell{0.98527\\(0.016)}  & \makecell{0.98728\\(0.014)}  & \makecell{0.98895\\(0.012)}  & \makecell{0.98998\\(0.011)}  & \makecell{0.99106\\(0.01)}  & \cellcolor[gray]{0.8}\makecell{\textbf{0.99167}\\\textbf{(0.009)}} \\ \hline 

            \end{tabular}%
        }
        \caption{FilmTrust}
        \label{tab:ndcg:ft}
        \end{subtable}
        \begin{subtable}[h]{1\textwidth}
        \resizebox{\textwidth}{!}{%
            \renewcommand{\arraystretch}{1.9}
            \begin{tabular}{|l|c|c|c|c|c|c|c|c|c|}
        \hline
Model \textbackslash Group Size & 2      & 3      & 4      & 5      & 6      & 7      & 8      & 9      & 10     \\ \hline
\makecell[l]{GMF IPA \\ GMF Avg} & \makecell{0.98898\\(0.014)}  & \cellcolor[gray]{0.8}\makecell{\textbf{0.99218}\\\textbf{(0.01)}}  & \cellcolor[gray]{0.8}\makecell{\textbf{0.99401}\\\textbf{(0.008)}}  & \makecell{0.9949\\(0.007)}  & \makecell{0.99571\\(0.006)}  & \cellcolor[gray]{0.8}\makecell{\textbf{0.99615}\\\textbf{(0.005)}}  & \makecell{0.99662\\(0.005)}  & \makecell{0.99679\\(0.005)}  & \makecell{0.99714\\(0.004)} \\ \hline 
GMF Expertise & \makecell{0.98893\\(0.014)}  & \makecell{0.99209\\(0.01)}  & \makecell{0.99383\\(0.008)}  & \makecell{0.99479\\(0.007)}  & \makecell{0.99566\\(0.006)}  & \makecell{0.99609\\(0.005)}  & \makecell{0.99658\\(0.005)}  & \makecell{0.99674\\(0.005)}  & \makecell{0.99708\\(0.004)} \\ \hline 
GMF Softmax & \cellcolor[gray]{0.8}\makecell{\textbf{0.98898}\\\textbf{(0.014)}}  & \makecell{0.99217\\(0.01)}  & \makecell{0.99399\\(0.008)}  & \cellcolor[gray]{0.8}\makecell{\textbf{0.99492}\\\textbf{(0.007)}}  & \cellcolor[gray]{0.8}\makecell{\textbf{0.99572}\\\textbf{(0.006)}}  & \makecell{0.99615\\(0.005)}  & \cellcolor[gray]{0.8}\makecell{\textbf{0.99664}\\\textbf{(0.005)}}  & \cellcolor[gray]{0.8}\makecell{\textbf{0.99679}\\\textbf{(0.005)}}  & \cellcolor[gray]{0.8}\makecell{\textbf{0.99715}\\\textbf{(0.004)}} \\ \hline 
MLP IPA & \makecell{0.98723\\(0.015)}  & \makecell{0.99062\\(0.011)}  & \makecell{0.99255\\(0.009)}  & \makecell{0.9936\\(0.008)}  & \makecell{0.99458\\(0.007)}  & \makecell{0.99507\\(0.006)}  & \makecell{0.99573\\(0.006)}  & \makecell{0.99591\\(0.005)}  & \makecell{0.99632\\(0.005)} \\ \hline 
\makecell[l]{MLP Avg \\ DeepGroup} & \makecell{0.9868\\(0.016)}  & \makecell{0.99023\\(0.012)}  & \makecell{0.99212\\(0.01)}  & \makecell{0.99307\\(0.008)}  & \makecell{0.99415\\(0.007)}  & \makecell{0.99459\\(0.007)}  & \makecell{0.99519\\(0.006)}  & \makecell{0.99537\\(0.006)}  & \makecell{0.99565\\(0.005)} \\ \hline 
MLP Expertise & \makecell{0.9867\\(0.016)}  & \makecell{0.99021\\(0.012)}  & \makecell{0.9921\\(0.01)}  & \makecell{0.99312\\(0.008)}  & \makecell{0.99414\\(0.007)}  & \makecell{0.99458\\(0.007)}  & \makecell{0.99521\\(0.006)}  & \makecell{0.99542\\(0.006)}  & \makecell{0.9957\\(0.005)} \\ \hline 
MLP Softmax & \makecell{0.98684\\(0.016)}  & \makecell{0.99028\\(0.012)}  & \makecell{0.99211\\(0.01)}  & \makecell{0.99308\\(0.008)}  & \makecell{0.99416\\(0.007)}  & \makecell{0.9946\\(0.007)}  & \makecell{0.99521\\(0.006)}  & \makecell{0.99539\\(0.006)}  & \makecell{0.99566\\(0.005)} \\ \hline 

            \end{tabular}%
        }
        \caption{MyAnimeList}
        \label{tab:ndcg:anime}
        \end{subtable}
    \end{table}

\FloatBarrier

\section{Discussion}

\label{sec:discussion}

The main goal of this research is to evaluate different aggregation techniques to make recommendations to groups. As shown in~\Cref{sec:experiments}, we can see different trends according to: a) the models used; b) the way group information is aggregated; c) the datasets on which they act; and d) the size of the groups.

Focusing on the models, we can see how \ac{MLP}, which has several hidden layers, obtains a lower MAE; however, \ac{GMF}, a simpler model, obtains a lower MSE. Although the \ac{MLP} model has great power in these types of problem, it seems to overfit, generating very good recommendations for some users in the group but bad ones for the rest, hence achieving higher MSE values. On the other hand, the \ac{GMF} model can obtain smaller maximum errors in each group, which means that no user in the group is badly affected by the recommendation. In the results, we can also observe how the models with higher maximum errors lead to a poorer order of items according to user preferences and obtain worse performance in the \ac{NDCG} metric.

Looking at the aggregation of users, we can see that the best performing user aggregation is the average, followed by a very similar performance by the Softmax. However, the use of expert user weighting without softmax produces worse results. Based on the results, we can observe that in models that do not use a deep architecture, with several hidden layers, the \ac{IPA} and \ac{GPA} strategies produce similar results when the aggregation function is a linear transformation of latent factors (\ac{GMF}). However, we can see how the non-linearity of \ac{MLP} produces different results between both two strategies.

Regarding the different datasets, we can see that there is a clear trend in the models that achieve the best results in complex datasets with a large number of users, items, and votes, such as Movilens or MyAnimeList, while in the FilmTrust dataset, with a smaller number of votes, there is no clear trend.

In terms of group size, as more users have the group, the probability of finding discrepancies between user preferences increases. Therefore, we can see how a larger group size leads to higher values in all error metrics.

\section{Conclusions and future work}

\label{sec:conclusions}

With the irruption of \ac{NN} in the world of \ac{CF}, the possibilities of their ability to find non-linear patterns within user preferences to generate better predictions are opening up. To use these systems to generate a recommendation for a group of users, we need to aggregate their preferences. As we have seen in this research, there are several key points at which aggregation can be performed. \ac{GPA} strategies do the aggregation before or inside the model, so they have the advantage of taking into account the preferences of the entire group and that a single feedforward step generates the prediction. Unlike the \ac{IPA} strategy, which requires multiple predictions for each user and performs the aggregations after the model. In this study, we have tested how different approaches to perform \ac{GPA} work in different datasets comparing different metrics.

As future work, there are two key factors to consider. First, in this research, the researchers have designed user aggregation techniques presented to the models; in future work, these functions will be explored by different machine learning models. The second key point is that in this work models perform a knowledge transfer from the model trained for individuals to make group predictions; it is shown that although the models have high performance (MAE improvement), they tend to overfit when working in groups (larger errors in group prediction leading to worse MSE). To solve this problem, future work will try to perform a specialization training stage for groups after individual training.

\section*{Declarations}

The authors of this paper declare that they have no conflict of interest.

This work has been co-funded by the \emph{Ministerio de Ciencia e Innovación} of Spain and the European Regional Development Fund (FEDER) under grants PID2019-106493RB-I00 (DL-CEMG) and the \emph{Comunidad de Madrid} under \emph{Convenio Plurianual} with the \emph{Universidad Politécnica de Madrid} in the actuation line of \emph{Programa de Excelencia para el Profesorado Universitario}.

\section*{Data availability statement}

The \texttt{MovieLens1M}, \texttt{FilmTrust} and \texttt{MyAnimeList} dataset along with the source code of the experiments that support the findings of this study are available in \texttt{neural-cf-for-groups} GitHub's repository [\url{https://github.com/KNODIS-Research-Group/neural-cf-for-groups}].

\bibliographystyle{plain}
\bibliography{sn-bibliography}

\end{document}